\documentclass[twocolumn,showpacs,preprintnumbers,draftclsnofoot]{revtex4}
\usepackage{amsmath}
\usepackage{amssymb}
\usepackage{graphicx}
\usepackage{dcolumn}
\usepackage{bm}
\usepackage{amssymb}
\usepackage{graphicx}
\usepackage{dcolumn}
\usepackage{bm}

\begin{document}

\title{ Floquet states of Valley-Polarized Metal with One-way Spin or Charge Transport in Zigzag Nanoribbons }
\author{Ma Luo\footnote{Corresponding author:luom28@mail.sysu.edu.cn} }
\affiliation{The State Key Laboratory of Optoelectronic Materials and Technologies \\
School of Physics\\
Sun Yat-Sen University, Guangzhou, 510275, P.R. China}

\begin{abstract}

Two-dimensional Floquet systems consisting of irradiated valley-polarized metal are investigated. For the corresponding static systems, we consider two graphene models of valley-polarized metal with either a staggered sublattice or uniform intrinsic spin-orbital coupling, whose Dirac point energies are different from the intrinsic Fermi level. If the frequency of irradiation is appropriately designed, the largest dynamical gap (first-order dynamical gap) opens around the intrinsic Fermi level. In the presence of the irradiation, two types of edge state appear at the zigzag edge of semi-infinite sheet with energy within the first-order dynamical gap: the Floquet edge states and the strongly localized edge states. In narrow zigzag nanoribbons, the Floquet edge states are gapped out by the finite size effect, and the strongly localized edge states remain gapless. As a result, the conducting channels of the nanoribbons consist of the strongly localized edge states. Under the first and second model, the strongly localized edge states carry one-way spin polarized and one-way charge current around the intrinsic Fermi level, respectively. Thus, the narrow zigzag nanoribbons of the first and second model have asymmetric spin and charge transmission rates, respectively. Quantum-transport calculations predict sizable pumped currents of charge and spin, which could be controlled by the Fermi level.

\end{abstract}

 \maketitle

\section{Introduction}

Floquet theory describes the quantum states of systems with a (temporally) periodically driven Hamiltonian, such as optically irradiated graphene \cite{Rodriguez08,Takashi09,Savelev11} or (temporally) periodically strained graphene \cite{Taboada17,Taboada171}. Novel types of topological phases have been predicted to appear in dynamical systems of 2D materials of the graphene family \cite{Inoue10,Kibis10,Calvo11,ShuTing16,Mukherjee18,Yunhua17,Ledwith18,HangLiu18,LongwenZhou18}. A motivation to study Floquet states in 2D materials is to construct topologically protected edge states \cite{Piskunow14,Usaj14,Claassen16,Tahir16,Puviani17,Hockendorf18} for electronic and spintronic applications. Optically irradiated graphene, which features low-energy excitations near the K and K$^{\prime}$ Dirac points of the Brillouin zone of a honeycomb lattice, has Floquet gaps of all order around energy levels $\varepsilon=\frac{1}{2}\hbar\Omega N$ \cite{PerezPiskunow15}, with $\Omega$ being the optical frequency and $N$ being an integer. The first-order gaps (i.e., the dynamical gaps induced by the first-order electron-photon coupling) lie around $\varepsilon=\pm\frac{1}{2}\hbar\Omega$, and the second-order gaps lie around $\varepsilon=0$. At the edge of the semi-infinite graphene, the topological edge states appear within the first-order and higher-order gaps. Because the first-order gap is larger than the higher-order gaps, we aim to engineer Floquet systems with a first-order gap lying around the intrinsic Fermi level ($\varepsilon=0$).

For graphene models with particle-hole symmetry, the Dirac points of both valleys lie at $\varepsilon=0$, so that the first-order gap of the corresponding Floquet systems always lies around $\varepsilon=\pm\frac{1}{2}\hbar\Omega$. The naive idea is to move the energy level of one Dirac point to $\varepsilon=\pm\frac{1}{2}\hbar\Omega$ such that the first-order gap is moved to $\varepsilon=0$. To maintain neutrality in the system, the energy level of another Dirac point is moved to $\varepsilon=\mp\frac{1}{2}\hbar\Omega$. If the pair of Dirac points lie in opposite valleys, the static system constitutes valley-polarized metal (VPM).

This article considers two graphene models of VPM. For the static systems, zigzag nanoribbons of the two models host strongly localized edge states (SLESs) at the zigzag edge, whose band structures connect the two valleys. The first model of VPM is graphene with staggered sublattice intrinsic spin-orbital coupling (SOC). This staggered sublattice intrinsic SOC is found in graphene with proximity coupling to the transition metal dichalcogenides (TMDCs) \cite{Gmitra15,Gmitra16,Morpurgo15}. The SLESs were recently proposed to be pseudohelical edge states (PHESs) \cite{Frank18}. The PHESs carry one-way spin polarized current around $\varepsilon=0$. The second model of VPM is graphene with uniform intrinsic SOC and an appropriate staggered sublattice on-site potential and magnetic exchange field. The model is more conveniently realized in silicene-like 2D materials \cite{Motohiko12} because of the large SOC and tunable staggered sublattice on-site potential. The SLESs carry one-way charge current around $\varepsilon=0$. For both models, the bulk states in the zigzag nanoribbons carry spin or charge current that offset the one-way currents of the SLESs. Thus, the quantum-transport of the zigzag nanoribbons are regular, i.e., the forward and backward transmission rates are the same, and the pumped currents of charge and spin are zero.

Under irradiation at an appropriate frequency, Floquet systems based on the two models of VPM have a first-order gap around $\varepsilon=0$ in the bulk band structures. The zigzag edge of a semi-infinite sheet hosts Floquet edge states with energy within the first-order gap. The Floquet edge states are weakly localized at the zigzag edge. In the narrow zigzag nanoribbons, the Floquet edge states are gapped out due to the finite size effect. On the other hand, the SLESs are negligibly influenced by both the irradiation and the finite size effect. Thus, the SLESs become the dominating conductive states around $\varepsilon=0$, thus determining the quantum-transport behavior of the nanoribbons. As a result, the irradiated zigzag nanoribbons of the two models have pumped current of spin and/or charge in addition to nonzero spin conductance.

The article is organized as follows: In Section II, the tight binding model of VPM on a honeycomb lattice with a time-dependent Hamiltonian is given, and the calculation methods for the Floquet band structure and conductance are presented. In Section III and IV, the numerical results of the Floquet state consisting of the first and second VPM model are presented, respectively. In Section IV, the conclusion is given.

\section{Model Hamiltonian and calculation method}

The tight binding model on a honeycomb lattice is a general model that describes graphene in addition to silicene and germanene. The effect of optical irradiation is described by time-dependent Peierls phases on nearest- and next-nearest-neighbor hopping. The time-dependent Hamiltonian is
\begin{eqnarray}
H&=&-\sum_{\langle i,j\rangle,s}\gamma_{ij}(t)c_{is}^{+}c_{js} \nonumber \\
&&+i\sum_{\langle\langle i,j\rangle\rangle,s,s'}\lambda_{I}^{i}(t)\nu_{ij}[\hat{s}_{z}]_{ss'}c_{is}^{+}c_{js'} \nonumber \\&&+\Delta\sum_{i,s}\xi_{i}c_{is}^{+}c_{is}+\lambda_{M}\sum_{i,s,s'}[\hat{s}_{z}]_{ss'}c_{is}^{+}c_{is'}
\label{hamiltonian}
\end{eqnarray}
where $i(j)$ is the index of the lattice site, $s(s')=\pm1$ is the spin index, $\gamma_{ij}(t)=\gamma_{0}f_{\langle i,j\rangle}(t)$ is the time-dependent nearest-neighbor hopping energy with $\gamma_{0}$ being the hopping parameter and $f_{\langle i,j\rangle}(t)$ being the time-dependent function, $c_{is}^{+}$($c_{is}$) is the creation (annihilation) operator of the electron at the i-th lattice site with spin s, $\hat{s}_{z}$ is the spin-z Pauli matrix, and $\nu_{ij}=\pm1$ represents clockwise or counterclockwise next-nearest-neighbor hopping. The summation with indices $\langle i,j\rangle$($\langle\langle i,j\rangle\rangle$) covers the nearest-neighbor (next-nearest-neighbor) lattice site. $\gamma_{0}$ is 2.8, 1.6, and 1.3 eV for graphene, silicene, and germanene, respectively. $\lambda_{I}^{i}(t)$ is equal to $\lambda_{I}^{A}f_{\langle\langle i,j\rangle\rangle}(t)$ and $\lambda_{I}^{B}f_{\langle\langle i,j\rangle\rangle}(t)$ for the A and B sublattice, respectively, which also include the time-dependent function $f_{\langle\langle i,j\rangle\rangle}(t)$. $\Delta$ is the strength of the staggered sublattice on-site potential, and $\xi_{i}=\pm1$ represents the A or B sublattice. In graphene, $\Delta$ can be induced by an h-BN \cite{Giovannetti07,CRDean10} or SiC \cite{SYZhou07} substrate; in silicene and germanene, $\Delta$ is induced by a vertical static electric field $E_{z}$. Because of the buckled structure of silicene and germanene, the A and B sublattice planes are separated by $2l$; thus, $\Delta=E_{z}l$. The exchange field $\lambda_{M}$ is induced by proximity with a ferromagnetic insulator. $\Delta$ and $\lambda_{M}$ are not time dependent. For the corresponding static systems, the first model of VPM has parameters $\lambda_{I}^{A}=-\lambda_{I}^{B}=\lambda_{I}$, $\Delta=0$ and $\lambda_{M}=0$, and the second model of VPM has parameters $\lambda_{I}^{A}=\lambda_{I}^{B}=\lambda_{I}$ and $\Delta=\lambda_{M}=3\sqrt{3}\lambda_{I}$. In addition to application to graphene-like 2D materials, the two models can be experimentally realized in cold atomic systems \cite{Goldman16,Aidelsburger13,Miyake13}.

In the presence of a normally incident optical field with the in-plane electric field being $\mathbf{E}=\hat{\mathbf{x}}E_{x}\sin(\Omega t)+\hat{\mathbf{y}}E_{y}\sin(\Omega t-\varphi)$, the time-dependent function of the nearest-neighbor hopping terms are given as
\begin{eqnarray}
&&f_{\langle i,j\rangle}(t)=exp\{i\frac{2\pi}{\Phi_{0}}\int_{\mathbf{r}_{i}}^{\mathbf{r}_{j}}\mathbf{A}(\mathbf{r},t)\cdot\mathbf{r}\} \\
&=& exp\{i\frac{2e}{\hbar\Omega}[E_{x}\hat{\mathbf{x}}\cdot\mathbf{r}_{ij}\cos(\Omega t)+E_{y}\hat{\mathbf{y}}\cdot\mathbf{r}_{ij}\cos(\Omega t-\varphi)]\} \nonumber
\end{eqnarray}
where $\Phi_{0}$ is the magnetic flux quantum, $\mathbf{r}_{ij}=\mathbf{r}_{j}-\mathbf{r}_{i}$, with $\mathbf{r}_{i}$ being the location of the i-th lattice site. The time-dependent function of the intrinsic SOC $f_{\langle\langle i,j\rangle\rangle}(t)$ has the same form. In this article, we consider only the circular polarized optical field with $E_{x}=E_{y}=E_{0}$ and $\varphi=\pi/2$. According to Floquet theory, the Floquet state is a time-periodic function written as $|\Psi_{\alpha}(t)\rangle=e^{-i\varepsilon_{\alpha}t/\hbar}\sum_{m=-\infty}^{+\infty}|u_{m}^{\alpha}\rangle e^{im\Omega t}$, with $\varepsilon_{\alpha}$ being the quasi-energy level of the $\alpha$-th eigenstate and $|u_{m}^{\alpha}\rangle$ being the corresponding eigenstate in the m-th Floquet replica. The Floquet states and the corresponding quasi-energy level are the solution of the equation
\begin{equation}
H_{F}|\Psi_{\alpha}(t)\rangle=\varepsilon_{\alpha}|\Psi_{\alpha}(t)\rangle
\end{equation}
where $H_{F}=H-i\hbar\frac{\partial}{\partial t}$ is the Floquet Hamiltonian. The time-dependent factor in the Hamiltonian can be expanded by the set of time-periodic functions $e^{im\Omega t}$ as
\begin{equation}
f_{\langle i,j\rangle}(t)=\sum_{m=-\infty}^{\infty}i^{m}f_{\langle i,j\rangle}^{m}e^{im\Omega t}e^{-im\varphi}
\end{equation}
with
\begin{equation}
f_{\langle i,j\rangle}^{m}=\sum_{m'=-\infty}^{\infty}
J_{m'}(\frac{2eE_{x}}{\hbar\Omega}\hat{\mathbf{x}}\cdot\mathbf{r}_{ij})
J_{m-m'}(\frac{2eE_{y}}{\hbar\Omega}\hat{\mathbf{y}}\cdot\mathbf{r}_{ij})e^{im'\varphi}
\end{equation}
where $J_{m}(x)$ is the m-th order first-type Bessel function of argument $x$. A similar expansion is applied to $f_{\langle\langle i,j\rangle\rangle}(t)$. In the direct product space (Sambe space), $\mathcal{R}\bigotimes\mathcal{T}$, with $\mathcal{R}$ being the Hilbert space and $\mathcal{T}$ being the space of the time-periodic function, the set of functions $\{|u_{m}^{\alpha}\rangle,m\in\mathbb{N}\}$ form the time-independent basis functions of the Floquet states. In this space, the Floquet Hamiltonian can be expressed as the time-independent block matrix $\mathcal{H}^{(m_1,m_2)}$, with $m_1$ and $m_2$ being the indices of replicas. The diagonal blocks $\mathcal{H}^{(m,m)}$ include three parts: the nearest- and next-nearest-neighbor hopping terms, whose hopping coefficients are renormalized by the factor $f_{\langle i,j\rangle}^{0}$ and $f_{\langle\langle i,j\rangle\rangle}^{0}$, respectively; the staggered sublattice on-site potential and magnetic exchange field; and the diagonal matrix $m\hbar\Omega\mathbf{I}$. The nondiagonal block includes the nearest- and next-nearest-neighbor hopping terms, whose hopping coefficients are renormalized by the factor $i^{m_{2}-m_{1}}f_{\langle i,j\rangle}^{m_{2}-m_{1}}e^{-i(m_{2}-m_{1})\varphi}$ and $i^{m_{2}-m_{1}}f_{\langle\langle i,j\rangle\rangle}^{m_{2}-m_{1}}e^{-i(m_{2}-m_{1})\varphi}$, respectively. The quasi-energy band structures of the bulk or nanoribbon version of the model can be obtained by diagonalization of the Floquet Hamiltonian with appropriate Bloch periodic boundary conditions. For the eigenstate of the $\alpha$-th quasi-energy, the weight of the static component (i.e., the $m=0$ replica) is given as $\langle u_{0}^{\alpha}|u_{0}^{\alpha}\rangle$. In the numerical calculation, the Floquet index m is truncated at a maximum value with $m\in[-m_{max},m_{max}]$. In general, calculation with larger $m_{max}$ gives more accurate results. If the investigation focuses on the first-order gap of the quasi-energy dispersion in the $m=0$ replica, $m_{max}=2$ gives sufficiently accurate results because all replicas with a single-photon energy difference from the first-order gap are included. Similarly, if the investigation focuses on the P-th order gap, $m_{max}=2P$ is required for sufficient accuracy.

The density of states of a semi-infinite sheet with a zigzag edge is calculated to visualize the dispersion of SLESs and Floquet edge states. Similar to the Floquet Hamiltonian, the Floquet Green's function can be expressed as a time-independent block matrix in the Sambe space: $\mathcal{G}^{(m_1,m_2)}$. The local density of states on the i-th lattice site is $-\frac{1}{\pi}\textit{Im}[\mathcal{G}^{(0,0)}_{ii}]$. The Floquet Green's function of the primitive cells at the left or right zigzag edge can be obtained by applying a recursive method \cite{Nardelli99,Calvo13,Lewenkopf13} to the Floquet Hamiltonian. Because the edge states are not completely localized at the terminal primitive cell, a backward recursive process is performed to calculate the Floquet Green's function of the internal primitive cells near the zigzag edge. The numerical results show that the SLESs are strongly localized at the terminal primitive cell, while the Floquet edge states are weakly localized. In our calculation, the density of states for each zigzag edge corresponds to the summation of the local density of states of fifty primitive cells near the terminal, given as
\begin{equation}
\rho_{left(right)}(\varepsilon,k_{y})=-\frac{1}{\pi}\sum_{i\in left(right)}{\textit{Im}[\mathcal{G}^{(0,0)}_{ii}(\varepsilon,k_{y})]} \label{DOSformula}
\end{equation}

\begin{figure}[tbp]
\scalebox{0.34}{\includegraphics{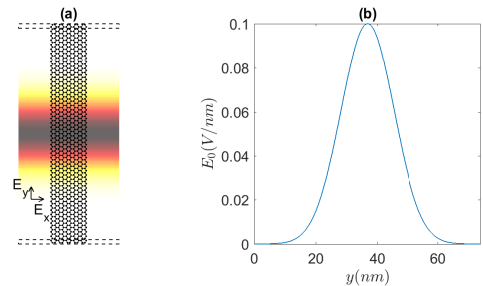}}
\caption{ (a) Structure of the transport calculation. The leads are within the dashed rectangle, which is not irradiated. Overlaid is the amplitude of the optical field, $E_{0}$. The function of $E_{0}$ versus the y coordinate is plotted in (b) and is a Gaussian function.  }
\label{fig_config}
\end{figure}

In practical circumstances, the transport is measured for a zigzag nanoribbon with a finite length that is connected to two leads \cite{Zhenghao11,Kitagawa11,FoaTorres14,Ying18}. The structure of the transport calculation is shown in Fig. \ref{fig_config}(a). The scattering region is the zigzag nanoribbon on the x-y plane with the width being 2.13 nm and the longitudinal length (along the y-axis) being 73.79 nm. The optical irradiation is restricted to the middle part of the scattering region. The amplitude of the optical field, $E_{0}$, is assumed to be uniform along the x-axis, and the Gaussian function along the y-axis is as shown in Fig. \ref{fig_config}(b). The leads are not irradiated by the optical field, so both leads are static systems, i.e., VPM. At the buffering unit cells (three unit cells in our calculation) between the leads and the scattering region, $E_{0}$ slowly increases from zero to a small value at the tails of the Gaussian function.

The Floquet Green's function of the zigzag nanoribbon in the scattering region is calculated by the recursive algorithm \cite{Nardelli99,Calvo13,Lewenkopf13}. For static systems, the transmission coefficient at energy level $\varepsilon$ from lead $L$ to lead $R$ is determined by the Landauer-Buttiker formula, $\mathcal{T}_{LR}(\varepsilon)=Tr[\Gamma_{L}(\varepsilon)\mathcal{G}_{LR}(\varepsilon)\Gamma_{R}(\varepsilon)\mathcal{G}_{LR}^{\dag}(\varepsilon)]$, with $\Gamma_{L(R)}(\varepsilon)$ being the decay width matrices of the $L$($R$) lead and $\mathcal{G}_{LR}(\varepsilon)$ being Green's function between the lattice sites that attach to the $L$ and $R$ leads. By contrast, for the Floquet systems, the transmission accompanied by the $m$-photon process (absorption for $m<0$ or emission for $m>0$) has transmission rate $\mathcal{T}_{LR}^{m}(\varepsilon)=Tr[\Gamma_{L}^{(m,m)}(\varepsilon)\mathcal{G}_{LR}^{(m,0)}(\varepsilon)\Gamma_{R}^{(0,0)}(\varepsilon)(\mathcal{G}_{LR}^{(m,0)})^{\dag}(\varepsilon)]$. The additional superscript $m$ is the index of the Floquet replicas, and $\mathcal{G}_{LR}^{(0,m)}$ is the $m$-th row $0$-th column block of the Floquet Green's function. Because the leads are not irradiated, the decay width matrices of each Floquet channel is given as $\Gamma_{L}^{(m,m)}(\varepsilon)=\Gamma_{L}(\varepsilon+m\hbar\Omega)$. The total forward (backward) transmission rate at energy $\varepsilon$ is $\mathcal{T}_{LR(RL)}(\varepsilon)=\sum_{m}{\mathcal{T}_{LR(RL)}^{m}(\varepsilon)}$ \cite{Calvo13,Ying18}. The transmission rates determine the time-average total current across the scattering region as \cite{FoaTorres14}
\begin{equation}
\mathcal{I}=\frac{2e}{h}\int{[\mathcal{T}_{LR}(\varepsilon)f_{L}(\varepsilon)-\mathcal{T}_{RL}(\varepsilon)f_{R}(\varepsilon)]d\varepsilon}
\end{equation}
where $f_{L(R)}(\varepsilon)$ is the Fermi-Dirac distribution at lead $L(R)$. In the presence of irradiation, the forward and backward transmission rates are different, so the current is possibly nonzero in zero bias, i.e., $f_{L}(\varepsilon)=f_{R}(\varepsilon)$. Assuming the zero temperature limit and linear order in the bias voltage $V$, the current is approximated as
\begin{eqnarray}
\mathcal{I}&=&\frac{2e}{h}\int_{-\infty}^{\varepsilon_{F}+\frac{eV}{2}}{\mathcal{T}_{LR}(\varepsilon)d\varepsilon}-\frac{2e}{h}\int_{-\infty}^{\varepsilon_{F}-\frac{eV}{2}}{\mathcal{T}_{RL}(\varepsilon)d\varepsilon} \nonumber \\
&\approx&G(\varepsilon_{F})V+\mathcal{I}^{p}(\varepsilon_{F})
\end{eqnarray}
where $\varepsilon_{F}$ is the Fermi level, $G(\varepsilon)=(2e^{2}/h)[\mathcal{T}_{LR}(\varepsilon)+\mathcal{T}_{RL}(\varepsilon)]/2$ is the differential conductance and $\mathcal{I}^{p}(\varepsilon_{F})=(2e/h)\int_{\infty}^{\varepsilon_{F}}{[\mathcal{T}_{LR}(\varepsilon)-\mathcal{T}_{RL}(\varepsilon)]d\varepsilon}$ is the pumped current. Because the transmission rates depend on the spin component, the differential conductance and the pumped current are spin-dependent, which are designated as $G_{\pm}$ and $\mathcal{I}^{p}_{\pm}$, respectively. One can define charge and spin differential conductances as $G_{C}=G_{+}+G_{-}$ and $G_{S}=G_{+}-G_{-}$, respectively, and the charge and spin pumped currents as $\mathcal{I}^{p}_{C}=\mathcal{I}^{p}_{+}+\mathcal{I}^{p}_{-}$ and $\mathcal{I}^{p}_{S}=\mathcal{I}^{p}_{+}-\mathcal{I}^{p}_{-}$, respectively.

\section{Graphene with staggered sublattice intrinsic SOC}

The first model of VPM is given by the Hamiltonian (\ref{hamiltonian}) with parameters $\lambda_{I}^{A}=-\lambda_{I}^{B}=\lambda_{I}$, $\Delta=0$ and $\lambda_{M}=0$. For a realistic heterostructure of graphene on TMDCs, the model should also include a staggered sublattice on-site potential and Rashba SOC. The heterostructure is an insulator instead of VPM. Engineering of the heterostructure, such as doping or positioning in proximity to another substrate, could offset the staggered sublattice on-site potential and Rashba SOC. We focus on the conceptual VPM model without Rashba SOC and with zero or small staggered sublattice on-site potential.

\subsection{Floquet states of the bulk}

The appropriate optical frequency depends on the model parameters and the optical parameters. If $\lambda_{I}$ is much smaller than $\gamma_{0}$, the low-energy excitations near the K and K$^{\prime}$ points of the Brillouin zone can be described by the Dirac fermion model with Hamiltonian
\begin{eqnarray}
H&=&\hbar v_{F}(\tau\sigma_{x}k_{x}+\sigma_{y}k_{y}) \\ \nonumber
&+&\frac{3\sqrt{3}}{2}[\lambda_{I}^{A}(\sigma_{z}+\sigma_{0})+\lambda_{I}^{B}(\sigma_{z}-\sigma_{0})]\tau s
\end{eqnarray}
where $\tau=\pm1$ stand for K or K$^{\prime}$ valleys. Because $\lambda_{I}^{A}=-\lambda_{I}^{B}=\lambda_{I}$, the intrinsic SOC terms become a constant potential $3\sqrt{3}\lambda_{I}\tau s$. The model has particle-hole-valley symmetry, i.e., the static band structure is symmetric under the simultaneous operation of particle-hole and K-K$^{\prime}$ valley exchanges. For the static systems, the energy levels of the Dirac points are $3\sqrt{3}\lambda_{I}s\tau$; all four Dirac fermion models are gapless. The intrinsic Fermi level cuts through all Dirac cones at finite energy, so the model corresponds to VPM. Although the particle-hole symmetry is broken, the band structure of each Dirac cone is symmetric about the energy level $3\sqrt{3}\lambda_{I}\tau s$. In the Floquet solution with $E_{0}=0$, the band structures of the $m$ replicas are obtained by adding $m\hbar\Omega$ to the static band structures. The crossings between the band structures of all replicas lie at energy $\varepsilon=3\sqrt{3}\lambda_{I}\tau s+\frac{1}{2}\hbar\Omega$. Therefore, the appropriate optical frequency is $\hbar\Omega=3\sqrt{3}\lambda_{I}\times2$. If $\lambda_{I}$ is not much smaller then $\gamma_{0}$, the static band structures at $\varepsilon=0$ significantly deviate from the Dirac fermion model such that the band structures are no longer symmetric about the energy level $3\sqrt{3}\lambda_{I}\tau s$. Thus, in the Floquet solution with $E_{0}=0$, the crossing between the band structures of all replicas is no longer aligned at energy $\varepsilon=3\sqrt{3}\lambda_{I}\tau s+\frac{1}{2}\hbar\Omega$. The optical frequency needs to be tuned around $\hbar\Omega=3\sqrt{3}\lambda_{I}\times2$ so that the crossings between the $m=0$ and $m=\pm1$ replicas align at $\varepsilon=0$. With $E_{0}\neq0$, the first-order gaps of the quasi-energy band structures lie around $\varepsilon=0$. However, the energy ranges of the first-order gaps in the two valleys are different from each other, so the global first-order gap is smaller than the first-order gap in each valley. Because the irradiation changes the hopping parameters in the diagonal blocks of the Floquet Hamiltonian by the factors $f_{\langle i,j\rangle}^{0}$ and $f_{\langle\langle i,j\rangle\rangle}^{0}$, the energy level of the Dirac points become dependent on $E_{0}$. As a result, for a given $E_{0}$, the frequency needs to be further tuned to maximize the global first-order gap.

\begin{figure}[tbp]
\scalebox{0.5}{\includegraphics{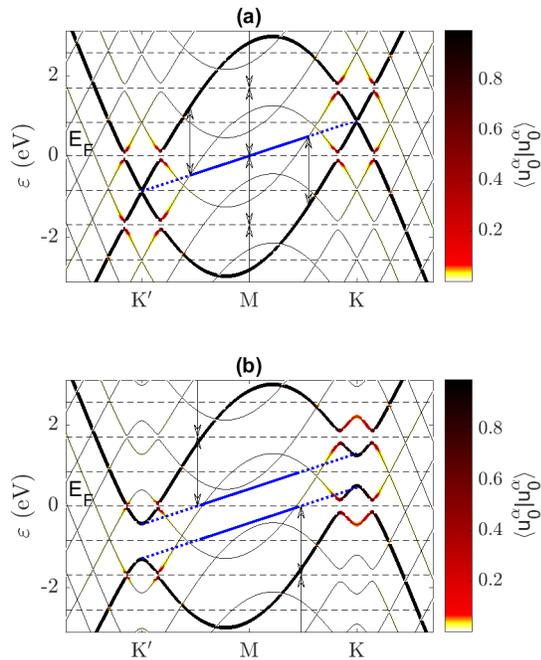}}
\caption{ The thin solid lines are the quasi-energy band structures of the spin-up electron plotted along the K-M-K$^{\prime}$ line in the Brillouin zone. The color scale on top of the band structures indicates the weight on the $m=0$ replica, $\langle u_{0}^{\alpha}|u_{0}^{\alpha}\rangle$. The model parameters are $\lambda_{I}^{A}=-\lambda_{I}^{B}=\lambda_{I}=0.06\gamma_{0}$ and $\lambda_{M}=0$; $\Delta=0$ in (a), and $\Delta=0.15\gamma_{0}$ in (b). The optical parameters are $E_{0}=0.3$ V/nm and $\hbar\Omega=3\sqrt{3}\lambda_{I}\times1.95$. The horizontal dashed lines indicate the energy $\frac{1}{2}\hbar\Omega N$. The thick (blue) straight lines indicate the energy levels of SLESs. In the dashed (solid) part, the difference between the energy levels of SLESs and bulk states is smaller (larger) than $\hbar\Omega$. The double arrows indicate the optical transition. }
\label{fig_bulk}
\end{figure}

The quasi-energy band structure of a spin-up electron in the bulk with parameters $\lambda_{I}=0.06\gamma_{0}$ and $E_{0}=0.3 V/nm$ is plotted in Fig. \ref{fig_bulk}(a). The band structure of the spin-down electron is obtained by mirroring the band structure of the spin-up electron about the M point. The optical frequency is tuned to $\hbar\Omega=3\sqrt{3}\lambda_{I}\times1.95$ so that the first-order gaps in K and K$^{\prime}$ valleys lie in the same energy range. Around energy $\varepsilon=0$, multiple side bands with small weight in the $m=0$ replica ($\langle u_{0}^{\alpha}|u_{0}^{\alpha}\rangle\ll1$) are gapless. As a result, the Floquet systems are not insulators. Thus, the topological property is not well defined for this Floquet system. In the additional presence of the staggered sublattice on-site potential, the local static gap of $2\Delta$ at the two Dirac points is opened. Assuming $\Delta=0.15\gamma_{0}$, the quasi-energy band structure is plotted in Fig. \ref{fig_bulk}(b). The first-order gaps in the two valleys both lie around $\varepsilon=0$. The gap size at the K (K$^{\prime}$) valley decreases (increases).

\subsection{Zigzag edge of semi-infinite sheet}

For a zigzag edge of semi-infinite sheet of the irradiated VPM, two types of edge states appear: the SLESs and Floquet edge states. The Floquet edge states appear in only the Floquet systems, with energy levels lying within the dynamical Floquet gaps. The SLESs appear in both the static and Floquet systems. With appropriate model parameters, the SLESs are negligibly impacted by the irradiation. For the static system of pristine graphene with $\lambda_{I}=0$, the bands of the SLESs are the zero-energy flat bands of the zigzag edge, which connect the two valleys. The SLESs are strongly localized at the terminal atom of one zigzag edge and weakly distributed among the other atoms in the same sublattice. With $\lambda_{I}\neq0$, the bands of the SLESs become nearly linearly dispersive with nonzero slope. For the graphene with staggered sublattice intrinsic SOC, the SLESs are PHESs. The PHESs with the same spin at the left and right zigzag edge travel along the same direction. By contrast, in the quantum spin Hall (QSH) model with uniform intrinsic SOC \cite{Kane2005}, the SLESs (referred to as helical edge states) with the same spin at the left and right zigzag edge travel along the opposite directions. Because the staggered sublattice intrinsic SOC does not induce band inversion, the PHESs appear in only the zigzag edge and do not appear in the armchair edge. The bulk band structures along the K-M-K$^{\prime}$ line in the Brillouin zone (Fig. \ref{fig_bulk}) correspond to the band edge of the bulk states in the semi-infinite sheet or nanoribbon with zigzag edge. One can plot the bands of the PHESs in the bulk band structure (the thick blue lines in Fig. \ref{fig_bulk}) and then estimate the optical coupling strength between the bulk states and the PHESs. If the difference between the energy levels of the PHESs and the bulk states is more than $\hbar\Omega$, the PHESs and the bulk states are negligibly coupled by higher-order photon transitions. The PHESs that satisfy this condition lie in the sections of the bands with solid blue lines in Fig. \ref{fig_bulk}. These sections would remain gapless in the Floquet systems. In contrast, the sections of the bands with dashed blue lines would be split by multiple Floquet gaps. For the model with $\Delta=0$ in Fig. \ref{fig_bulk}(a), the bands of the PHESs around $\varepsilon=0$ would remain gapless, and the PHESs would be the dominating conductive states. In contrast, for the model with $\Delta=0.15\gamma_0$ in Fig. \ref{fig_bulk}(b), the bands of the PHESs around $\varepsilon=0$ would be split, and the PHESs would not be the dominating conductive states. Thus, a small or vanishing staggered sublattice on-site potential is preferred. In the rest of this section, $\Delta=0$ is assumed.

\begin{figure}[tbp]
\scalebox{0.6}{\includegraphics{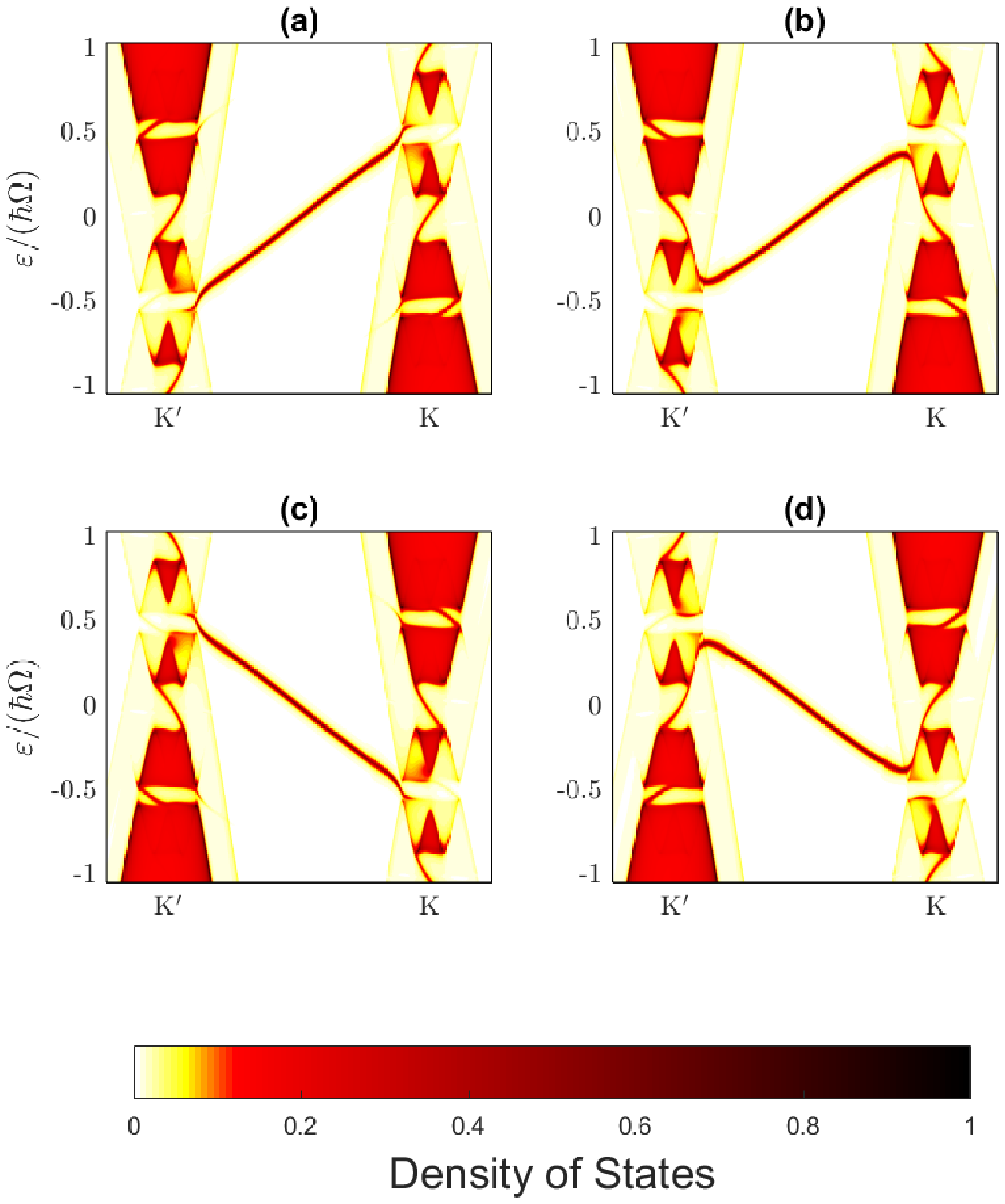}}
\caption{ Density of states for the zigzag edge of the semi-infinite sheet. $\rho_{left}(\varepsilon,k_{y})$ and $\rho_{right}(\varepsilon,k_{y})$ of the spin-up states are plotted in (a) and (b), respectively; those of the spin-down states are plotted in (c) and (d), respectively. The model parameters are $\lambda_{I}^{A}=-\lambda_{I}^{B}=\lambda_{I}=0.02\gamma_{0}$ and $\Delta=\lambda_{M}=0$. The optical parameters are $\hbar\Omega=3\sqrt{3}\lambda_{I}\times1.99$ and $E_{0}=0.1$ V/nm. The color scale is normalized to one. }
\label{fig_DOS}
\end{figure}

For the model with $\lambda_{I}=0.02\gamma_{0}$, $\hbar\Omega=3\sqrt{3}\lambda_{I}\times1.99$ and $E_{0}=0.1 V/nm$, the density of states of the spin-up (down) electron at the left and right zigzag edge is plotted in Fig. \ref{fig_DOS}(a) and (b) ((c) and (d)), respectively. The figures show the distribution of bulk states, Floquet edge states, PHESs and side bands in the ($\varepsilon$,$k_{y}$) space. The numerical results confirm that the PHESs have linear dispersive bands with wavenumber between the K and K$^{\prime}$ points; the PHESs with spin up (down) at both zigzag edges travel along the forward (backward) direction. Thus, the PHESs at the two zigzag edges carry spin currents along the same direction. The first-order gaps of the bulk states around $\varepsilon=0$ are approximately 0.1 eV. Within the first-order gap, the Floquet helical edge states (FHESs) appear. For the same zigzag edge and the same spin, the FHESs in the K and K$^{\prime}$ valleys travel along the same direction. For the same zigzag edge and the opposite spin, the FHESs travel along the opposite directions so that the FHESs carry spin current along the zigzag edge. For the opposite zigzag edge, the spin currents carried by the FHESs flow opposite to each other. The side bands have a small density of states around $\varepsilon=0$, which also contributes to the transport along the zigzag edge.

\subsection{Spin transport of the zigzag nanoribbon}

\begin{figure}[tbp]
\scalebox{0.46}{\includegraphics{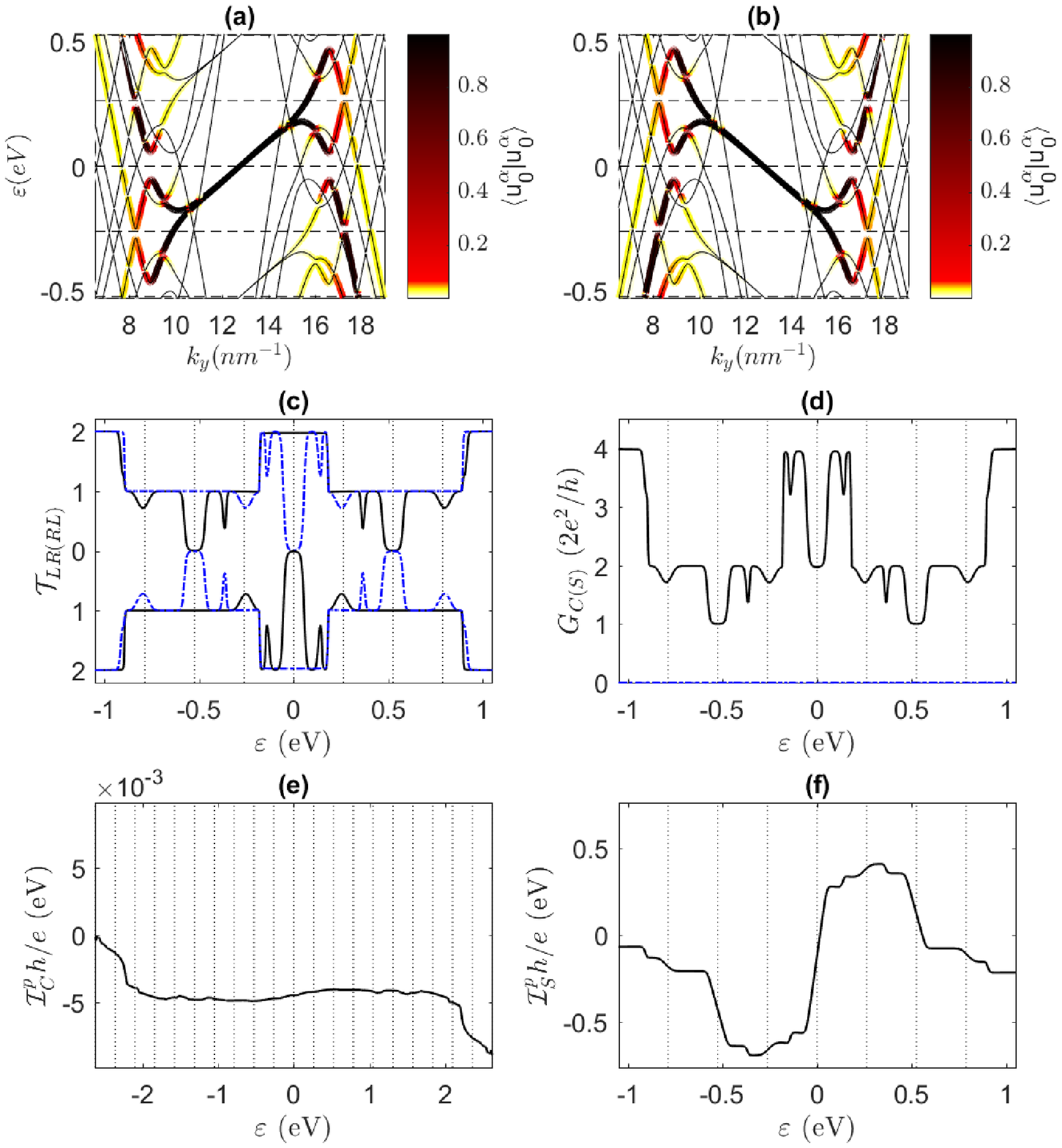}}
\caption{ The quasi-energy band structures of spin-up and spin-down electrons of the zigzag nanoribbon are shown in (a) and (b), respectively. The color scale on top of the band structures indicates the weight on the $m=0$ replica, $\langle u_{0}^{\alpha}|u_{0}^{\alpha}\rangle$. The width of the zigzag nanoribbon is 2.13 nm. The model and optical parameters are the same as in Fig. \ref{fig_DOS} except that the optical frequency is changed to $\hbar\Omega=3\sqrt{3}\lambda_{I}\times1.8$. The horizontal dashed lines indicate the energy $\frac{1}{2}\hbar\Omega N$. The forward (backward) transmission rates of the spin-up and spin-down electrons versus energy are plotted in (c) as solid black (dashed blue) lines in the upper and lower scale of the y-axis, respectively. The charge and spin differential conductances are plotted in (d) as solid black and dashed blue lines, respectively. The charge and spin pumped current are plotted in (e) and (f), respectively. The vertical dotted lines in (c-f) indicate the energy $\frac{1}{2}\hbar\Omega N$.  }
\label{fig_ribbon}
\end{figure}

In the narrow zigzag nanoribbons, the appropriate optical frequency depends on the nanoribbon width. The band structures of the bulk states and FHESs significantly deviate from those in the zigzag edge of the semi-infinite sheet (as shown in Fig. \ref{fig_DOS}) due to the finite size effect. This effect mixes the bulk states and the FHESs at the two zigzag edges, forming the nanoribbon mixed states. The optical frequency needs to be tuned again so that the first-order gaps of the nanoribbon mixed states in the two band valleys lie in the same energy range. On the other hand, the band structure of the PHESs with wavenumbers between the two band valleys are only minimally impacted by the finite size effect due to strong localization at the zigzag terminal. A zigzag nanoribbon with a width of 2.13 nm is studied as an example. For the model with $\lambda_{I}=0.02\gamma_{0}$ and $E_{0}=0.1 V/nm$, the optical frequency is tuned to $\hbar\Omega=3\sqrt{3}\lambda_{I}\times1.8$. The quasi-energy band structures of spin-up and spin-down electrons are plotted in Fig. \ref{fig_ribbon}(a) and (b), respectively. Within the energy range of the first-order gaps, the side bands have only a small weight on the $m=0$ replica, so the PHESs become the dominating conductive bands. The forward (backward) quantized transmission rate of the PHESs with spin-up (down) electrons is $2$, whereas the backward (forward) transmission rate of the PHESs with spin-up (down) electrons is $0$. As a result, the zigzag nanoribbon exhibits one-way spin polarized transmission. Because of the presence of the conductive side bands, the difference between the forward and backward transmission rate of the same spin is slightly smaller than $2$.

To confirm the one-way spin polarized transport, quantum-transport calculations were performed for the finite irradiated zigzag nanoribbon with the structure shown in Fig. \ref{fig_config}(a). The forward (backward) transmission rate $\mathcal{T}_{LR(RL)}$ versus energy are plotted in Fig. \ref{fig_ribbon}(c). For better visualization, $\mathcal{T}_{LR(RL)}$ of the spin-up and spin-down electrons is plotted in the upper and lower scale of the y-axis, respectively. The spin-up $\mathcal{T}_{LR(RL)}$ is equal to the spin-down $\mathcal{T}_{RL(LR)}$, so the differential conductances of the spin-up and spin-down electrons are the same. Thus, the spin differential conductance is zero. The charge and spin differential conductances are plotted in Fig. \ref{fig_ribbon}(d). The charge differential conductance consists of multiple plateaus with dips. The charge and spin pumped current are plotted in Fig. \ref{fig_ribbon}(e) and (f), respectively. Although the charge conductance is finite, the charge pumped current is nearly zero. The magnitude of this current monotonically increases as the Fermi level rises. On the other hand, the spin pumped current is finite, and its direction is controlled by the Fermi level. If $\varepsilon_{F}=0$, the spin pumped current is small but nonzero. If $\varepsilon_{F}$ changes within the first-order Floquet gap, the spin pumped current rapidly changes. In the absence of optical irradiation, both $\mathcal{T}_{LR}$ and $\mathcal{T}_{RL}$ are the same, so the spin pumped current vanishes. Therefore, one-way spin transport is controlled by the presence of the optical irradiation.

The optical parameters for experimental implementation of the Floquet system are discussed here. For a bulk or semi-infinite sheet, we assume that the graphene is irradiated by a normally incident Gaussian beam. If the width of the beam waist is larger than the wavelength, the optical field in the middle of the Gaussian beam can be idealized as a plane wave. We designate $w_{0}=w_{1}\lambda$ as the width of the beam waist with $\lambda=2\pi c/\Omega$ being the wavelength and $w_{1}\geq1$. The power of the Gaussian beam is $P_{0}=\frac{\pi|E_{0}|^{2}w_{0}^{2}}{4\mathrm{Z}_{0}}$ with $\mathrm{Z}_{0}=\sqrt{\frac{\mu_{0}\mu_{r}}{\varepsilon_{0}\varepsilon_{r}}}$ being the impedance of the background media. The first-order gap can be estimated by a first-order perturbation method as $\eta\hbar\Omega=\frac{ev_{F}E_{0}}{\Omega}$\cite{Usaj14} with $\eta<0.5$ and $v_{F}\approx c/330$. Thus, the power of the Gaussian beam is
\begin{equation}
P_{0}=\frac{\pi c^{2}\hbar^{2}\Omega^{2}w_{1}^{2}\eta^{2}}{\mathrm{Z}_{0}ev_{F}^{2}}\approx(30\hbar\Omega w_{1}\eta)^{2} [W]
\end{equation}
For the model with parameters in Fig. \ref{fig_DOS} and \ref{fig_ribbon}, assuming that $w_{1}=1$ and $\eta=0.2$, we have $P_0\approx12$ W. For the system in Fig. \ref{fig_ribbon}, the optical field pattern has subwavelength size. Plasmonic devices, such as a metallic tip or plasmon cavity \cite{Miyazaki06}, can be used to focus the Gaussian beam into a subwavelength field pattern. The local electric field is enhanced by the geometry factor $F$. Thus, the required power of the laser beam is reduced by a factor of $\sqrt{F}$.

\section{Graphene with uniform intrinsic SOC }

The second VPM model is given by the Hamiltonian (\ref{hamiltonian}) with parameters $\lambda_{I}^{A}=\lambda_{I}^{B}=\lambda_{I}$ and $\Delta=\lambda_{M}=3\sqrt{3}\lambda_{I}$. This model is more conveniently realized in the 2D staggered semiconductors silicene, germanene, stanene, and plumbene \cite{Alessandro17}.

\subsection{Floquet states of the bulk}

\begin{figure}[tbp]
\scalebox{0.5}{\includegraphics{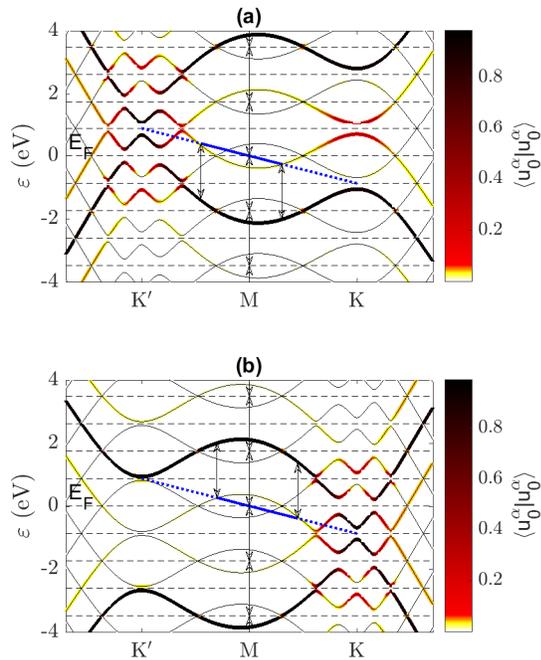}}
\caption{ The quasi-energy band structure of the spin-up and spin-down electrons in (a) and (b), respectively, are plotted along the K-M-K$^{\prime}$ line in the Brillouin zone. The color scale on top of the band structures indicates the weight on the $m=0$ replica, $\langle u_{0}^{\alpha}|u_{0}^{\alpha}\rangle$. The model parameters are $\lambda_{I}^{A}=\lambda_{I}^{B}=\lambda_{I}=0.06\gamma_{0}$ and $\Delta=\lambda_{M}=3\sqrt{3}\lambda_{I}$; the optical parameters are $E_{0}=0.9$ V/nm and $\hbar\Omega=2\lambda_{M}$. The horizontal dashed lines indicate the energy $\frac{1}{2}\hbar\Omega N$. The thick (blue) straight lines indicate the energy levels of SLESs. In the dashed (solid) part, the difference between the energy levels of SLESs and bulk states is smaller (larger) than $\hbar\Omega$. The double arrows indicate the optical transition. }
\label{fig_bulk_ME}
\end{figure}

The appropriate optical frequency depends on the model parameters but not on the optical parameters. The low-energy excitation near the K and K$^{\prime}$ points of the Brillouin zone can be described by the Dirac fermion model, whose Hamiltonian is
\begin{eqnarray}
H&=&\hbar v_{F}(\tau\sigma_{x}k_{x}+\sigma_{y}k_{y}) \\
&+&3\sqrt{3}\lambda_{I}\sigma_{z}\tau s+\Delta\sigma_{z}+\lambda_{M}\sigma_{0}s \nonumber
\end{eqnarray}
The model has particle-hole-valley-spin symmetry, i.e., the static band structure is symmetric under the simultaneous operation of particle-hole, K-K$^{\prime}$ valley and spin-up and spin-down exchanges. For the static systems, the energy
levels of the Dirac points are $\lambda_{M}s$. Two Dirac fermion models (spin up in the K$^{\prime}$ valley and spin down in the K valley) are gapless; the other two Dirac fermion models have a gap as large as $4\lambda_{M}$. The intrinsic Fermi level cuts through the two gapless Dirac cones, which have opposite spin and lie in opposite valleys. Thus, the VPM exhibits spin-valley locking. With sizable intrinsic SOC, the band structures at $\varepsilon=0$ deviate from the linear dispersion of the Dirac cones. In addition, the band structure of spin-$s$ electron is symmetric about the energy level $\lambda_{M}s$ throughout the whole Brillouin zone. As a result, the appropriate optical frequency is exactly $\hbar\Omega=\lambda_{M}\times2$. In the presence of irradiation with this frequency, the first-order gaps and all higher-order gaps of the Floquet quasi-energy band structures lie around $\varepsilon=0$. The Floquet systems are insulators. The quasi-energy band structures of the Floquet system with model parameters $\lambda_{I}=0.06\gamma_{0}$ and optical parameters $E_{0}=0.9 V/nm$ and $\hbar\Omega=2\lambda_{M}$ are plotted in Fig. \ref{fig_bulk_ME}.

Because the Floquet systems are insulators, their topological properties can be well defined. One can define the Chern number $\mathcal{C}_{\alpha}$ of the $\alpha$-th quasi-energy band \cite{PerezPiskunow15,Kitagawa10,Rudner13} as
\begin{equation}
\mathcal{C}_{\alpha}=\frac{1}{2\pi}\int_{BZ}d^{2}k\mathcal{B}_{\alpha} \label{chernNum}
\end{equation}
where $\mathcal{B}_{\alpha}$ is the Berry curvature of the $\alpha$-th quasi-energy band. The integral covers the whole Brillouin zone. For the Floquet systems, $\mathcal{B}_{\alpha}$ is defined as
\begin{eqnarray}
&&\mathcal{B}_{\alpha}(\mathbf{k})=\label{berryC} \\
&&-\sum_{\alpha^{\prime}\ne\alpha}\frac{2\emph{Im}\langle\Psi_{\alpha}(\mathbf{k},t)|v_{x}|\Psi_{\alpha^{\prime}}(\mathbf{k},t)\rangle\langle\Psi_{\alpha^{\prime}}(\mathbf{k},t)|v_{y}|\Psi_{\alpha}(\mathbf{k},t)\rangle}{(\varepsilon_{\alpha}-\varepsilon_{\alpha^{\prime}})^{2}} \nonumber
\end{eqnarray}
where $v_{x(y)}=\nabla_{k_{x}(k_{y})}H_{F}(\mathbf{k})$ is the velocity operator. The Berry curvature can be evaluated at any fixed time. The truncation of the Floquet replicas should satisfy the condition $m_{max}>6\gamma_{0}/(\hbar\Omega)$, which ensures that the quasi-energy bands within the static bandwidth $6\gamma_{0}$ include all relevant crossing between different replicas. For spin $s$, the lowest $2m_{max}+1-s$ quasi-energy bands are occupied. Summation of the Chern numbers of all occupied bands gives the winding number at the intrinsic Fermi level. The winding number yields the number of Floquet edge states across the intrinsic Fermi level. However, for the Floquet systems in Fig. \ref{fig_bulk_ME}, numerical calculation of Eq. (\ref{chernNum}) is challenging because $\mathcal{B}_{\alpha}(\mathbf{k})$ has sharp peaks at the higher-order dynamical gap. For the Floquet systems with large optical intensity ($E_{0}>20$ V/nm), the winding numbers of each spin vary between $-3$ and $1$. In some ranges of $E_{0}$, the winding numbers of the two spins are different. In the remaining part of this article, the discussion focuses on the Floquet systems with more realistic optical intensity, although the corresponding winding number was not calculated.

\subsection{Zigzag edge of semi-infinite sheet}

For a zigzag edge of semi-infinite sheet of the irradiated VPM, the SLESs appear in both the static and Floquet systems. Similar to the analysis in Section IIIB and Fig. \ref{fig_bulk}, the bands of the SLESs are plotted in the bulk band structure in Fig. \ref{fig_bulk_ME} as thick blue lines. For each spin, only one of the two SLESs has a band structure that crosses the energy $\varepsilon=0$. For the model parameters in Fig. \ref{fig_bulk_ME}, the bands of SLESs around $\varepsilon=0$ remain gapless in the Floquet systems. This feature is valid for the realistic systems with smaller $\lambda_{I}$.

\begin{figure}[tbp]
\scalebox{0.6}{\includegraphics{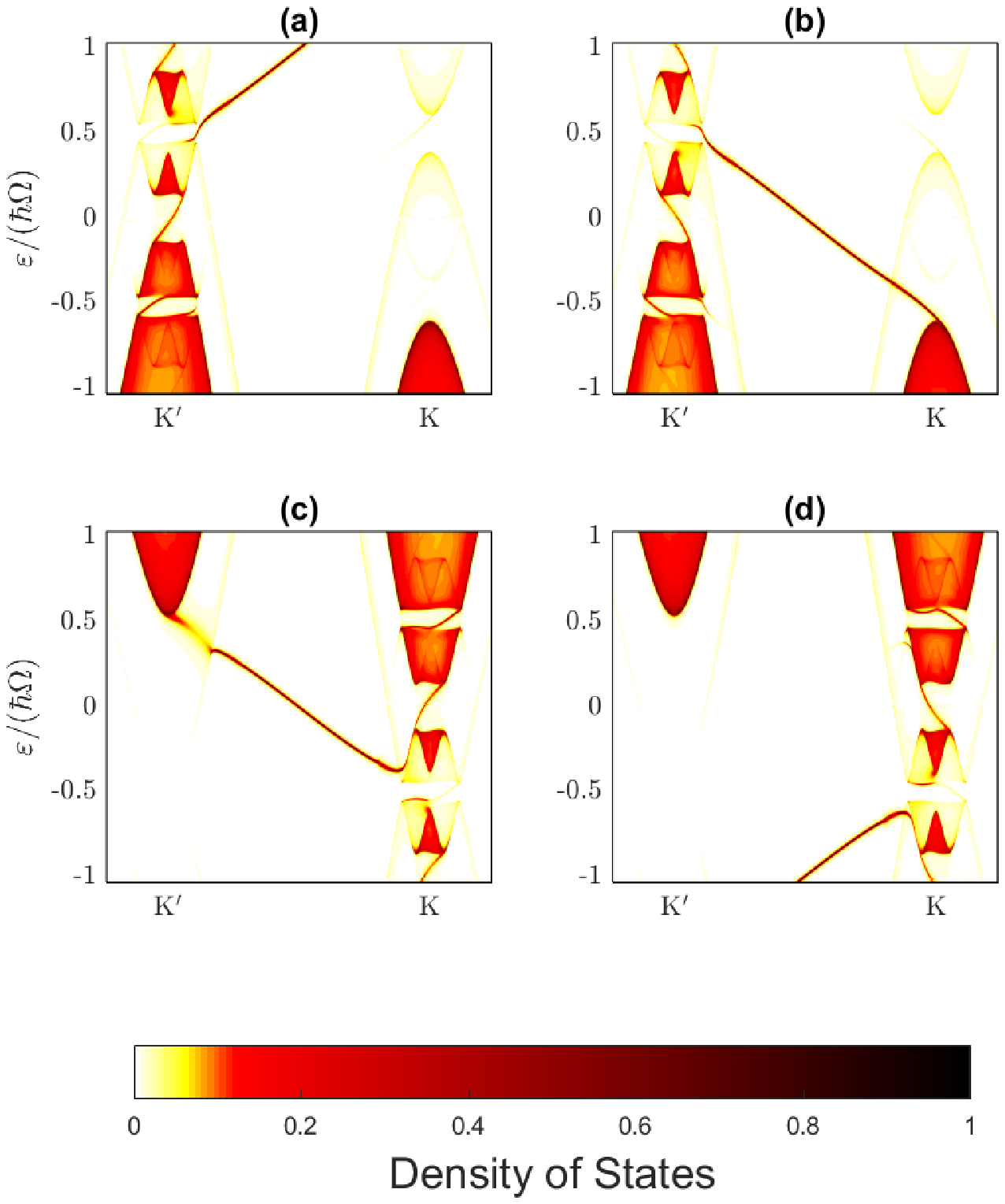}}
\caption{ Density of states for the zigzag edge of the semi-infinite sheet. $\rho_{left}(\varepsilon,k_{y})$ and $\rho_{right}(\varepsilon,k_{y})$ for the spin-up states are plotted in (a) and (b), respectively; those of the spin-down states are plotted in (c) and (d), respectively. The model parameters are $\lambda_{I}^{A}=\lambda_{I}^{B}=\lambda_{I}=0.02\gamma_{0}$ and $\Delta=\lambda_{M}=3\sqrt{3}\lambda_{I}$. The optical parameters are $\hbar\Omega=\lambda_{M}\times2$ and $E_{0}=0.1$ V/nm. The color scale is normalized to one. }
\label{fig_DOS_ME}
\end{figure}

For the model with $\lambda_{I}=0.02\gamma_{0}$, $\hbar\Omega=\lambda_{M}\times2$ and $E_{0}=0.1 V/nm$, the density of states of the spin-up (down) electron at the left and right zigzag edge is plotted in Fig. \ref{fig_DOS_ME}(a) and (b) ((c) and (d)), respectively. The numerical results confirm that the bands of the two SLESs cross $\varepsilon=0$. The two SLESs have opposite spin, are localized at opposite zigzag terminals, and travel along the same direction. The Floquet edge states with bands within the first-order gap around $\varepsilon=0$ are designated Floquet chiral edge states (FCESs). Different from the FHESs in the previous model, the FCESs carry charge currents at the zigzag edge. The charge currents at the opposite zigzag edges flow opposite to each other.

\subsection{Charge and spin transport of the zigzag nanoribbon}

\begin{figure}[tbp]
\scalebox{0.46}{\includegraphics{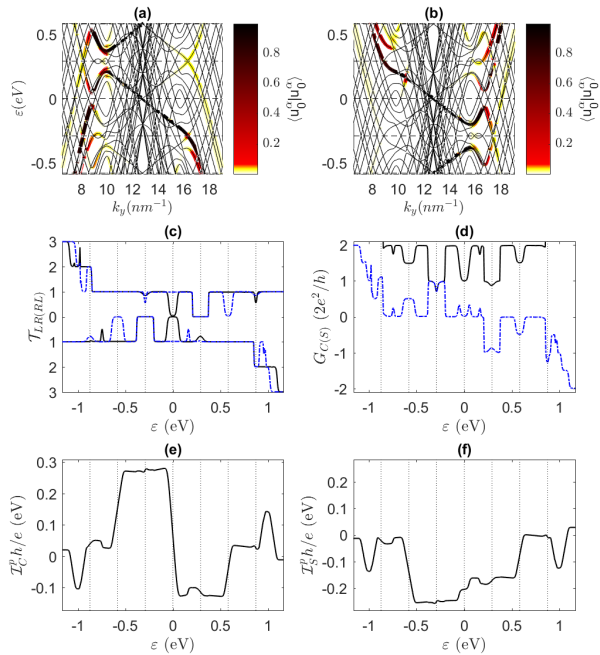}}
\caption{ The quasi-energy band structures of the spin-up and spin-down electrons of the zigzag nanoribbon are plotted in (a) and (b), respectively. The color scale on top of the band structures indicates the weight on the $m=0$ replica, $\langle u_{0}^{\alpha}|u_{0}^{\alpha}\rangle$. The width of the zigzag nanoribbon is 2.13 nm. The model and optical parameters are the same as in Fig. \ref{fig_DOS_ME}. The horizontal dashed lines indicate the energy $\frac{1}{2}\hbar\Omega N$. The forward (backward) transmission rates of the spin-up and spin-down electrons versus energy are plotted in (c) as solid black (dashed blue) lines in the upper and lower scale of the y-axis, respectively. The charge and spin differential conductances are plotted in (d) as solid black and dashed blue lines, respectively. The charge and spin pumped current are plotted in (e) and (f), respectively. The vertical dotted lines in (c-f) indicate the energy $\frac{1}{2}\hbar\Omega N$.  }
\label{fig_ribbon_ME}
\end{figure}

In the narrow zigzag nanoribbons, the appropriate optical frequency ($\hbar\Omega=\lambda_{M}\times2$) is not altered by the finite size effect because the static band structure of spin-$s$ is symmetric to the energy level $\lambda_{M}s$. For the zigzag nanoribbon with a width of 2.13 nm, $\lambda_{I}=0.02\gamma_{0}$, $E_{0}=0.1 V/nm$ and $\hbar\Omega=\lambda_{M}\times2$, the quasi-energy band structures of the spin-up and spin-down electrons are plotted in Fig. \ref{fig_ribbon_ME}(a) and (b), respectively. The dominating conductive state around $\varepsilon=0$ corresponds to the SLESs that carry one-way charge current. As a result, with the Fermi level around $\varepsilon=0$, the forward and backward quantized transmission rate is expected to be $0$ and $2$, respectively. The large band gaps of the spin-up and spin-down band structures around the energy $\frac{1}{2}\hbar\Omega$($-\frac{1}{2}\hbar\Omega$) arise not from optical irradiation but from the finite size effect. The optical irradiation induces the Floquet side bands within these gaps.

Similar to the analysis in the previous model, the transport of the zigzag nanoribbon with finite length and a restricted irradiated region is calculated. The forward and backward transmission rates versus energy are plotted in Fig. \ref{fig_ribbon_ME}(c). At $\varepsilon=0$, the only conductive states are the SLESs that travel along the backward direction, so the forward transmission rate should be zero. However, the numerical result of the forward transmission rate at $\varepsilon=0$ is $0.03$ because of tunneling through the irradiated region. For the longer scattering region with a 900-unit cell, the forward transmission rate is reduced to $10^{-6}$. The charge and spin differential conductances are plotted in Fig. \ref{fig_ribbon_ME}(d). The differential conductances of the spin-up and spin-down electrons are different, so the spin differential conductance is nonzero. The plateaus of quantized spin differential conductance around $\pm\frac{1}{2}\hbar\Omega$ are due to the finite size effect rather than the optical irradiation. The dips in these plateaus are due to the excitation of the side bands by the irradiation. The charge and spin pumped currents are plotted in Fig. \ref{fig_ribbon_ME}(e) and (f), respectively. Because the first-order gaps at $\varepsilon=0$ and $\varepsilon=\hbar\Omega$($\varepsilon=-\hbar\Omega$) induce large differences between the forward and backward transmission rates of the spin-up and spin-down electrons, the pumped current of each spin rapidly changes as the Fermi level sweeps through these gaps. As the Fermi level sweeps through $\varepsilon=0$, the charge pumped current changes rapidly and changes its sign, while the spin pumped current changes only gradually. As $\varepsilon_{F}=16$ meV, the charge pumped current is zero, so the pumped current is pure spin current.

\section{conclusion}

In conclusion, the Floquet systems of optically irradiated VPM consisting of 2D graphene-like materials are investigated. Two graphene models of VPM are considered. For the corresponding static systems, the SLESs of the first (second) model carry one-way spin polarized (one-way charge) current. By choosing the appropriate optical frequency and strength, the Floquet systems feature a first-order dynamical gap around the intrinsic Fermi level, which gaps out the conductive bulk states in bulk, semi-infinite sheet and nanoribbon configurations. At the zigzag edge of the semi-infinite sheet, the conductive states are the SLESs, Floquet edge states and side bands. In narrow zigzag nanoribbons, the conductive states are SLESs and side bands. The transport of the side bands is negligible; thus, the transport of the narrow zigzag nanoribbons is determined almost entirely by the properties of the SLESs. As a result, the one-way spin or charge transport is optically induced. By sweeping the Fermi level within the first-order gap, the direction and magnitude of charge or spin pumped current can be controlled.

\begin{acknowledgments}
The project is supported by the National Natural Science Foundation of China (grant no.
11704419), the National Basic Research
Program of China (grant no. 2013CB933601), and the National Key Research and Development Project of China
(grant no. 2016YFA0202001).
\end{acknowledgments}

\clearpage

\end{document}